# Adsorption and Intermolecular Interaction of Cobalt Phthalocyanine on CoO(111) Ultrathin Films: An STM and DFT Study

*Tobias Schmitt†, Pascal Ferstl†, Lutz Hammer†, M. Alexander Schneider†\*, and Josef Redinger‡*

†Solid State Physics, Friedrich-Alexander-Universität Erlangen-Nürnberg (FAU), Staudtstr. 7, 91058 Erlangen, Germany

‡Institute of Applied Physics, Vienna University of Technology, Wiedner Hauptstr. 8-10/134 , A-1040 Vienna, Austria

\*alexander.schneider@physik.uni-erlangen.de

We investigate the adsorption of cobalt phthalocyanine (CoPc) molecules on a thin layer of cobalt oxide grown on Ir(100). To that end we compare the results of low-temperature scanning tunneling microscopy (STM) with those of ab-initio density functional theory (DFT) calculations and reveal the adsorption geometry. We find that the CoPc molecules lie flat on the surface and that their central cobalt atom forms a chemical bond to a substrate oxygen ion. However, this bond contributes only modestly to the adsorption energy, while van-der-Waals forces dominate the potential landscape and determine to a large extend the geometry as well as the distortion of substrate and molecule. Furthermore they lead to attractive molecule–molecule interactions at higher molecular coverages. The DFT calculations predict energetic positions of the molecular orbitals which are compared to scanning tunneling spectroscopy (STS) measurements.

## 1. Introduction

Phthalocyanines are one of the most popular molecules in surface science since they offer a wide spectrum of applications like gas sensing[1], molecular electronics[2,3] or dye-sensitized solar cells[4-6] where often oxide substrates are used. Many of the applications require a degree of molecular ordering at the interface and thus a detailed understanding of the molecule–substrate and molecule–molecule interaction is necessary in order to create the desired structures by molecular self-assembly. So far phthalocyanines have been extensively studied on metal surfaces over the last 25 years[7–17] and in recent years an increasing number of studies also used metal oxides as substrates[18–30] with the main focus on $TiO_2$[27–30] but to our knowledge no investigations on (ultrathin) cobalt oxide (CoO) were published yet. Also to what extent chemical bonds between the phthalocyanine molecule and the substrate contribute to the adsorption energy has not been established.

The aim of our paper is therefore to clarify the interaction mechanisms between a phthalocyanine molecule and the substrate and also between molecules in supramolecular structures by a detailed comparison between experimental STM data and DFT calculations. We employ the Generalized-Gradient Approximation (GGA) to DFT augmented by the inclusion of van-der-Waals (vdW) interactions for the adsorption of cobalt phthalocyanine (CoPc) on a cobalt oxide (111) thin film substrate. In that respect, the chosen molecule can be considered as a representative of a class of molecules (e.g. also porphyrin molecules) where coordinated metal ions and aromatic backbones can interact with an oxidic substrate.

CoO shows a large variety of different phases[31], as well as intriguing catalytic[32] and magnetic[33,34] properties and hence is interesting for fundamental research and technological applications. Using thin films allows us to investigate the system with STM that offers a high spatial resolution. However, such ultrathin films often differ from bulk termination due to the oxide/substrate interface which can lead to heavy distortions within the oxide layer. In the present paper we discuss the adsorption of CoPc on a quasi-hexagonal CoO(111) film that exhibits up to 1 Å height differences in the positions of the oxygen atoms. Due to the prevalence of vdW forces the adsorption is only moderately influenced by this.

## 2. Methods

Cobalt oxide was prepared on an Ir(100) single crystal under ultrahigh vacuum conditions (base pressure $1 \cdot 10^{-10}$ mbar). First the iridium crystal was cleaned by several cycles of Ne sputtering (3 keV) and subsequent annealing in oxygen at 1300 K. The unreconstructed Ir(100)-(1×1) surface was prepared following standard recipes[38,39].

The thin one bilayer CoO(111) films were prepared by depositing 0.9 ML of cobalt onto the substrate kept at 323 K. Subsequent post oxidizing at 523 K leads to a well ordered film with low defect density[36].

Cobalt(II)phthalocyanine molecules (Sigma-Aldrich, 97% purity) were evaporated from a Knudsen cell held at T = 723 K after thorough degassing. The molecular coverage was varied and is given in the text whereas one monolayer is referred to a fully covered layer of flat-lying molecules (0.9 molecules/nm$^2$). The surfaces were investigated with a homebuilt UHV-STM operating at 80 K, for preparations with a molecular coverage close to 1 ML a different UHV-STM operating at room temperature (RHK-UHV-300) was used. All topographies were acquired in the constant current mode with the bias voltage applied to the sample.

The Vienna Ab-initio Simulation Package (VASP) employing the projector augmented-wave formalism was used to perform the ab-initio calculations[40]. Since a proper treatment of dispersion effects is crucial for a correct description of CoPc adsorption on the ultra thin CoO(111) surface the optB86b functional[41,42] was used. This functional improves the GGA[43] by accounting for the vdW interactions (vdW-DF) according to the original formalism of Dion[44]. Adsorption was modelled putting a single CoPc molecule on a suitable three times enlarged c(10×2) CoO(111)/Ir supercell. Our previous experience on the substrate system[36,45] allowed for a restriction of the Ir thickness to 3 layers of the asymmetric slab to reliably model the energetics and adsorption geometries. Generally, an energy cutoff of 400 eV and a 3×3×1 k-point grid of Monckhorst-Pack[46] type were used to ensure electronic convergence. To analyze the importance of the vdW interactions the calculations were compared to standard GGA calculations using PBE functionals.[47] All atomic coordinates were relaxed until the residual forces were below 0.01 eV/Å. In order to maintain consistency with our previous studies[36,45] the calculations were performed for a substrate Ir lattice constant of 3.88 Å, corresponding to the bulk value determined for a PBE functional. Typical bond lengths for the CoPc molecule are 1.91 Å for the Co–N bond, 1.09 Å for the C–H bond 1.40 Å and 1.45 Å for the C–C bonds in the benzene and imide ring, respectively. The constant-current STM simulations rely on the Tersoff–Hamann approach to Bardeen tunneling[48].

## 3. Results and Discussion

The properties of one bilayer (1 BL) of CoO(111) on Ir(100) have been investigated by experiment and theory in great detail[35,36], and thus the atomic structure is known with high accuracy and reproduced by the theoretical approach employed here. The 1 BL film is a distorted hexagonal CoO(111) plane that is perfectly row matched to the substrate lattice along [011] of the Ir crystal. Along the orthogonal [01$\bar{1}$] direction the film adopts no strict registry with the substrate, instead it forms a one-dimensional moiré.[36] The moiré structure is evidenced by the strong buckling of the oxygen atoms (~1 Å) out of the plane of the film. Whether an oxygen atom is in the Co plane or buckled out of it depends on its registry with the Ir substrate: O atoms that laterally sit near Ir top sites are within the Co plane ($O_{low}$, labeled L1, L2, L3) while those that end up near Ir bridge sites are pushed 1 Å above the Co plane ($O_{high}$, labeled H1, H2,) c.f. Fig.1. The $O_{high}$ site is a Co threefold hollow site with no direct coupling to the Ir substrate. In the course of our work presented here we became aware of the fact that small variations of the amount of Co evaporated modifies the lattice parameter of the CoO film and hence the observed length of the moiré. This can be traced by a careful analysis of the LEED pattern but also by the appearance of the CoO surface in STM. Here, the $O_{high}$ atoms are imaged and along [01$\bar{1}$] different sequences of high- and low-lying oxygen atoms are observed (Fig. 1). The true film structure lies between a commensurate c(10×2) and c(8×2), depending on the Co coverage the film is closer to one or the other. In the following we use the two limiting commensurate structures to characterize the moiré length of the oxide film. In our theoretical treatment the unit cell is approximated by a c(10×2) structure that reproduces the buckling and bond lengths of the clean substrate.[36] For the determination of adsorption geometries we neglect the slightly improved fit to the experimental substrate geometry by a DFT+U treatment (U − J = 1 eV) as discussed in Ref. 36.

When single CoPc molecules are deposited at room temperature they are intact and adsorb flat lying on the CoO film analogous to the situation on other oxide films[21–26,29,30]. On a CoO film closer to a c(10×2) structure, the molecules end up in the dominant configuration shown in Figure 1a. By comparison with the atomically resolved lattice of the CoO we arrive at the model shown in Figure 1b. Similarly, on a film approximated by a c(8×2) cell we find the configuration shown in Figure 1c and d. Although it was expected that the central Co ion of the CoPc molecule binds to an oxygen atom of the surface, it is quite surprising to find that the molecule is either exclusively centered on an $O_{low}$ on the c(10×2) or on an $O_{high}$ on the c(8×2) structure. Considering the rotational degree of freedom ("orientation") of the molecules we find two distinct, symmetrically equivalent orientations where the molecular mirror axis (defined as the central line connecting two opposite isoindole units) is rotated by ±27±5° with respect to the [011] direction, independent of the actual CoO moiré length. This low-symmetry configuration is due to the complex corrugated van-der-Waals potential between surface and molecule, as will be shown below.

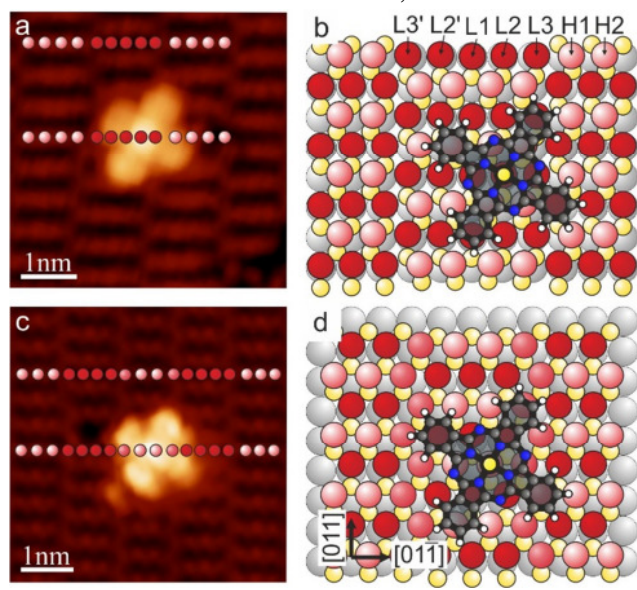

Figure 1. STM images of CoPc molecules adsorbed on a 1 bilayer thick CoO film and their corresponding suggested models. (a) CoPc adsorbed on a CoO film approximated by a c(10×2) structure where the molecule is centered on low-lying oxygen atoms (U = 0.5 V, I = 0.4 nA) and (b) corresponding model. (c) and (d) the same for CoPc adsorbed on a CoO film approximated by c(8×2) structure where the molecule is centered to high-lying oxygen atoms (U = 0.4 V, I = 0.5 nA). In both cases the molecules are oriented -27±5° with respect to the [011] direction. Substrate's color coded ball model: light grey spheres – iridium atoms, yellow spheres – cobalt atoms, light/dark red spheres – high-/low-lying oxygen atoms.

At higher molecular coverages we observe a subtle difference in the self-assembly depending on the lattice parameter (i.e. moiré length) of the CoO substrate. For approximately 1 ML CoPc on the more relaxed c(10×2) substrate all molecules in dense assemblies change their orientation and are rotated by 45°±10° with respect to [011] (Figure 2a,b). Due to the dense coverage we could not determine the adsorption sites of specific molecules experimentally. However, a local $\begin{pmatrix} 5 & -1 \\ 0 & 10 \end{pmatrix}$ superstructure with respect to the Ir(100) surface containing two molecules is consistent with the STM images and we propose for the densely packed film the arrangement of the molecules as shown in Figure 2c. In this superstructure neighboring molecules along [011] must alternately occupy $O_{high}$ and $O_{low}$ sites. Whereas the alignment along $\vec{a}$ is fixed, the rows of molecules parallel to $\vec{a}$ can shift with respect to their neighboring rows. Hence, the long-range order is disturbed and the molecular film consists of only small domains. The structures shown in Figure 2 are also rationalized by the results from our vdW-DFT calculations discussed below. We determine a smallest intermolecular Co-Co distance of 1.40±0.05 nm along the vector $\vec{a}$ of Figure 2c. Distances larger than that by integer numbers of the Ir lattice parameter are found in places where the molecular layer is not perfectly ordered. The shortest Co–Co distance between molecules on different adsorption sites is 1.43±0.10 nm.

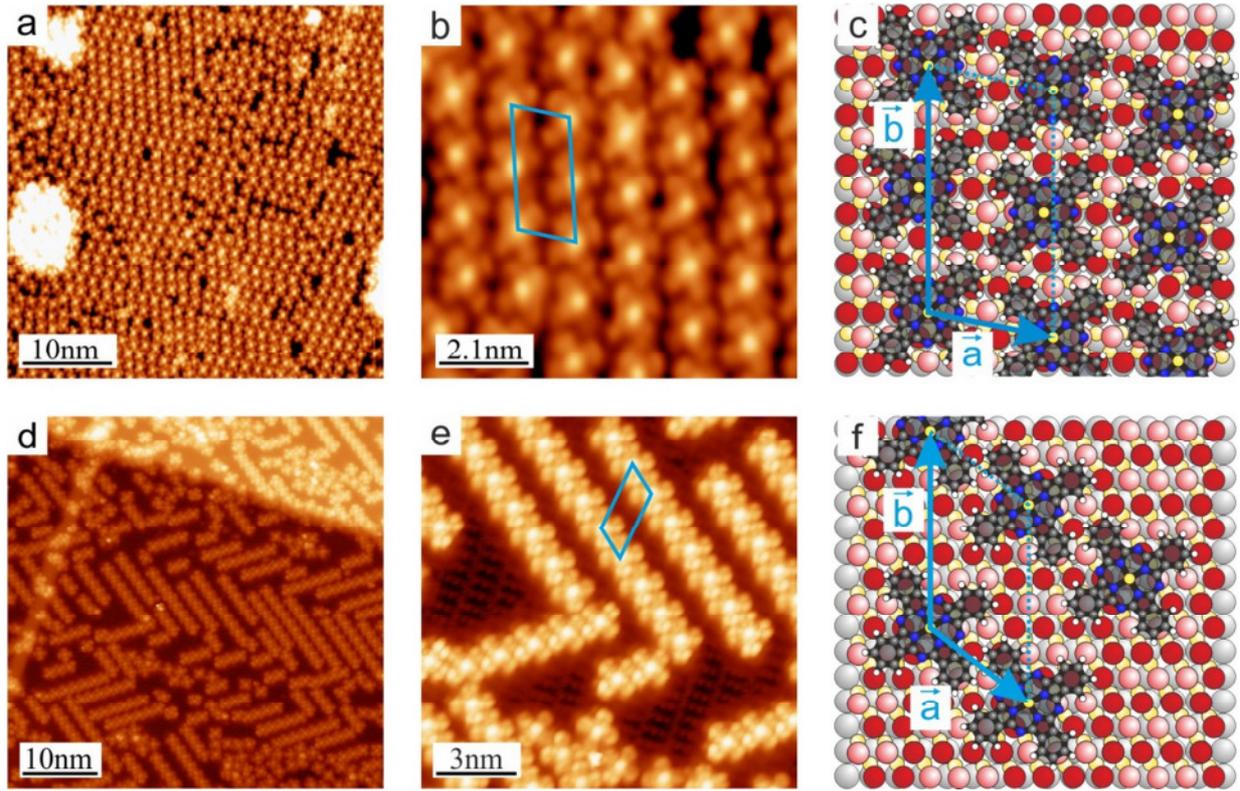

Figure 2. (a,b) STM images of ~1ML CoPc on the relaxed CoO c(10×2) phase. The molecules are densely packed in a locally ordered structure. (U = 0.4 V, I = 2.9 nA) (d,e) STM images of 0.7 ML CoPc adsorbed on a strained CoO c(8×2) phase. CoPcs form chains with a length of 2 – 10 molecules if the sample is kept at RT during adsorption ((d) U = 0.8 V, I = 0.1 nA; (e) U = 0.1 V, I = 0.1 nA). The suggested ball model for the ordered molecular configurations along with the superstructure cell is shown in (c) for the densely packed and (f) for the chain structure.

A different scenario prevails if 0.7 ML CoPc are adsorbed on a c(8×2) CoO film at RT. Linear, one-dimensional, self-assembled chains are formed (Figure 2d,e) and are oriented ±55±10° with respect to [011] that consist of approximately 2 to 10 molecules. This indicates that there is a significant molecule-molecule interaction. The molecules are mobile enough to diffuse while the sample is kept at RT during preparation but the thermal energy is not enough to rearrange already formed chains into longer ones. As in the case of low coverage the CoPc are oriented 27±5° with respect to the [011] direction and the centers are only located above one of the $O_{high}$ in the c(8×2) unit cell. The observed intermolecular distance of a = 1.33±0.05 nm is the smallest possible molecule-molecule distance if the same adsorption sites and 27° rotated molecules are assumed. In dense agglomerations molecular chains are separated by a distance of b = 2.20±0.05 nm or b = 2.70±0.05 nm measured along [011] which corresponds to 8 or 10 Ir unit cells. This is the minimum distance if identical adsorption sites are assumed since for inter-chain distances smaller than 8 $a_{Ir}$ the molecules would touch each other, hence the assembly is dictated by the available adsorption sites and the minimal distance the molecules may have between them. We identify preferred molecular superstructures of $\begin{pmatrix} 4 & -3 \\ 0 & 8 \end{pmatrix}$ or $\begin{pmatrix} 4 & -3 \\ 0 & 10 \end{pmatrix}$ and propose the structural model shown in Figure 2f.

To improve our understanding of the subtle effects of the molecular interaction with the CoO layer we have performed vdW-DFT calculations of molecules on the c(10×2) layer in a $\begin{pmatrix} 5 & -1 \\ 0 & 6 \end{pmatrix}$ superstructure with an orientation corresponding to the model of Figure 2c but neglecting the switch in the molecular adsorption site from $O_{low}$ to $O_{high}$ and thereby reducing the size of the unit cell. The results of this calculation are summarized in Figure 3. First of all: without vdW correction the calculations would not result in a bound state of the molecule. With vdW correction the adsorption energy is $E_{ads} = -(E_{total} - E_{CoPc} - E_{CoO/Ir}) = 3.6$ eV. Placing the molecular Co ion on an O bridge site reduces $E_{ads}$ by ~500 meV. Of the five symmetrically inequivalent oxygen top sites two are clearly unfavorable. Of the remaining three the adsorption energy on the $O_{high}$ (H1) site is the highest followed by an 18 meV decrease on the nearby $O_{low}$ (L3) site (Figure 3 a). This difference vanishes and $E_{ads}$ is increased by 40 meV if the molecular orientation is changed to 40° instead of 45° which is the energetically lowest configuration (well within the limits of the experimental results). The (almost) equivalence of the two sites for the 40°/45° oriented molecules is surprising because although the Co-O bond length is nearly the same for the two configurations (H1: 2.06 Å, L3: 2.14 Å) the $O_{low}$ atom at site L3 is pulled out from its original position by 0.9 Å while it is only 0.1 Å for the H1 site. Such a large change in the height of the $O_{low}$ atom at site L3 might seem very surprising at first sight, but one has to remember the special structural properties of the CoO thin film as discussed in Ref. 36. Depending on the local interaction with the substrate, strong attraction and repulsion points for both cobalt and oxygen atoms lead to severe distortions in the film. The site L3 has been identified as such a repulsion point and hence already a small change in the local environment may lead to a considerable height change with only small energetic cost. By comparison the Co ion of the molecule is vertically shifted 0.2 Å towards the surface out of the plane of the macrocycle. On the scale of the buckled substrate the molecule itself remains essentially flat (Figure 3 b), the next smallest distance between atoms of the molecule and of the substrate is found to occur for the N–O distance of 2.73 Å which shows that there is no other chemical bond involved than that of the central ion with the substrate O atoms. From calculations of free base 2H-Pc adsorption the contribution of the Co–O bond may be estimated to be 0.49 eV or 14 % of the total adsorption energy. The finding of a Co–O bond is in contrast to the analysis of a related system in Ref. 23 where a dipolar interaction is proposed for the interaction of MgPc with a thin FeO film.

The forces between surface and molecule induce further relaxations of the substrate. In Figure 3 c and d we plot the displacement of the surface oxygen atoms upon adsorption of CoPc and 2H-Pc respectively. Displacements of the O atoms below the molecule of up to 0.2 Å are calculated. By comparing the displacement patterns it becomes apparent that the response of oxygen atoms lying below the outer phenyl rings is independent of the strong displacement induced by the molecular Co–O bond and can therefore be considered as the response of the substrate to the repulsive interaction with the electron density of the π-system. Also, no special role of the nitrogen atoms of the molecule can be discerned. In consequence, the slight distortion of the molecular backbone of 0.2 Å from an ideal flat configuration essentially follows the topology of the underlying substrate.

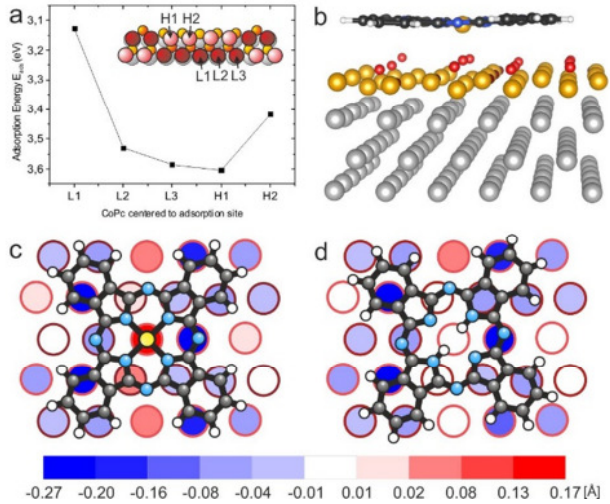

Figure 3. Results of DFT+vdW calculation of the proposed model of Fig. 2 c. (a) Adsorption energy as function of the position of the central Co ion of CoPc for the five inequivalent O top sites within the c(10×2) unit cell. Connecting lines are guides to the eye. Not shown is the ~500meV barrier for bridge site adsorption. (b) Side view along $[01\bar{1}]$ showing the molecule adsorbed on H1. The oxygen atom closest to the Co ion of the molecule is pulled outwards from an undisturbed position and the molecule is slightly distorted from a flat configuration. (c) Color coded representation of the distortion of the oxygen lattice in the CoO layer upon CoPc adsorption on site H1. Blue colors represent atoms pushed into the surface, red colors those with an increase of the distance to the Ir substrate. Perimeter colors correspond to those of the model in (a) for $O_{high}$ and $O_{low}$. (d) Same as (c) for the adsorption of a 2H-Pc on H1 with a slightly different optimal orientation. Color scale given is valid for both (c) and (d).

In order to get further insight into the intermolecular interactions, especially those causing the linear molecular arrangements on the c(8×2) strained substrate, DFT energy calculations of free CoPc molecules in the observed superstructures were performed (Figure 4). While this is a severe simplification it allows us to check for intrinsic (i.e. not surface-mediated) molecular interactions as function of distance and rotational angle. While for PBE almost no attractive interactions can be found, calculations including vdW corrections show a well-defined energy minimum with a gain of up to 134 meV per molecule. Thus the main contribution to this energy must come from the polarizability of nearby phenyl rings.[49] The molecule-molecule vdW interaction energy is of the order of expected diffusion barriers. Hence, at room temperature the molecules can test different configurations and form the ordered structure in optimal geometry that is not simply a closed-packed arrangement. For the configuration equivalent to the $\begin{pmatrix} 4 & -3 \\ 0 & 8 \end{pmatrix}$ supercell (with 27.5° orientation of the molecules within the cell) a calculated optimal CoPc-CoPc distance of 1.40 nm was found which is somewhat larger than the measured distance of 1.33±0.05 nm (Figure 2 e). The experimentally found close-to-optimal intermolecular distance is maintained on the c(8×2) substrate by minimizing the molecular interaction energy while the molecule-substrate interaction forces the molecule onto a particular site as outlined above. Since this combination is only optimal in one direction, the molecules will preferentially align in chains that are, however, easily interrupted by arrangements in the symmetrically equivalent chain directions. From these observations we would exclude any significant substrate mediated interaction that causes the end position of a chain to be a preferential nucleation site for a diffusing molecule[50].

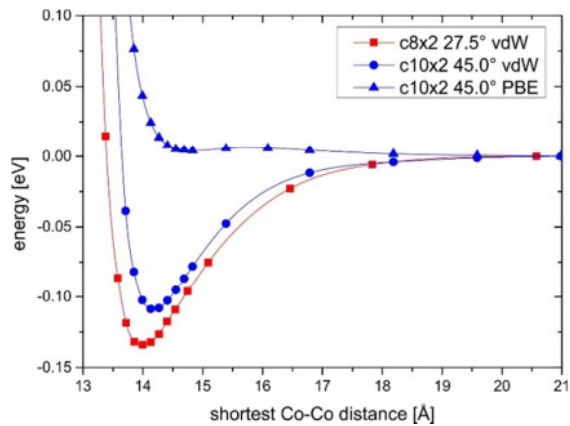

Figure 4. To test whether the molecule-molecule interaction that causes the ordered structure at higher CoPc coverage is substrate-mediated, the interaction energy of free molecules in a planar arrangement was calculated. For simulating the situation on the c(8×2) CoO(111) substrate a supercell corresponding to that of the model in Fig. 2 f was used with 27.5° molecular orientation whereas for the c(10×2) the CoPc were oriented at 45°. The results show an attractive van-der-Waals interaction with the minimum energy distances located close to those of the c(8×2) and c(10×2) structure.

A similar gain in energy is found for the empty lattice approximation to the dense molecular structure on the relaxed c(10×2) substrate. Since the molecules are rotated with respect to the previous case the energy minimum occurs at a slightly different molecular distance of 1.41 nm. This value agrees very well with the intermolecular distances found in experiment. Hence we can conclude that the molecules arrange themselves in a dense packing pattern dictated by the optimization of intermolecular vdW interactions while centering to O atoms. This overwrites easily the slight preference of the H1 site over the L3 site which differ only by 18 meV (cf. Figure 3 a) and causes the alternating adsorption sites.

We will now turn to the electronic structure of CoPc adsorbed on 1 bilayer CoO. For that we analyze the appearance of the molecules in STM topographies of different bias voltages. We find that the appearance is independent of the molecule's adsorption site, orientation and CoO moiré length. From the experimental data (Figure 5 a) taken on linear molecular chains of the c(8×2) we find that in case of a filled state image at negative bias voltage the molecule's appearance is dominated by the carbon macrocycle with a well visible intramolecular structure and additionally a minor contribution of the central Co ion. For intermediate voltages up to approximately +1 V the internal contrast of the molecular backbone diminishes and the Co center dominates the molecular appearance. For voltages larger than ~ +1.5 V (empty states) a different intramolecular orbital structure reappears. In Figure 5 b we show the corresponding simulated STM images from our vdW-DFT calculations of the $\begin{pmatrix} 5 & -4 \\ 0 & 6 \end{pmatrix}$ superstructure on a c(10×2), which additionally employ a DFT+U (U–J = 1 eV) treatment[51] of the central molecular Co atom. Within these approximations the observed patterns are well reproduced only the energetic position of the HOMO deviates (see discussion below). Therefore we can conclude that the basic contributions of the various molecular orbitals to the tunneling current and the spatial structure of these orbitals are well captured by our calculations.

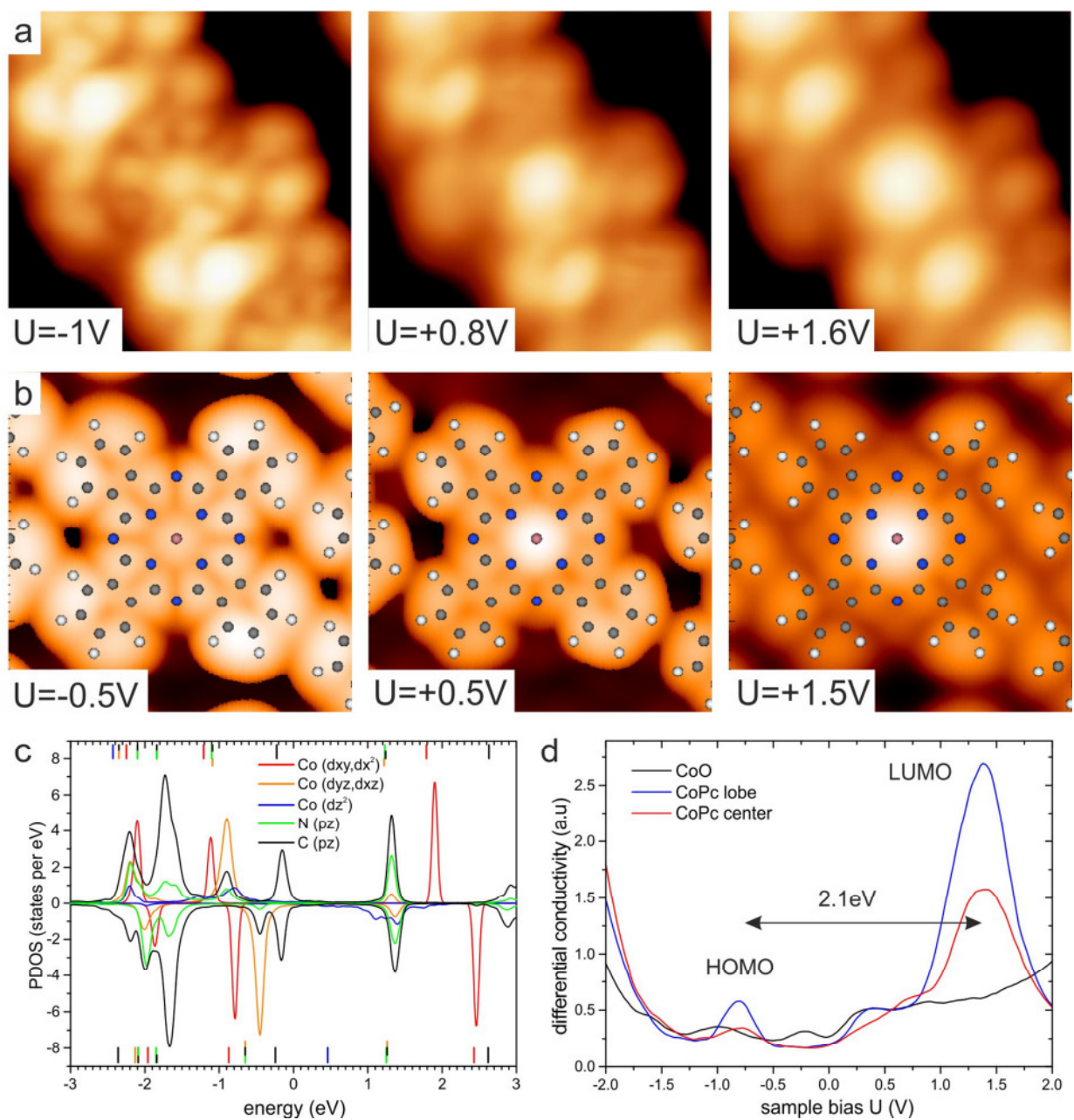

Figure 5. Submolecular structure of CoPc adsorbed on CoO 1BL. (a) shows the measured STM images and (b) the simulated counterparts at different bias voltages U. (c) The calculated projected density of states (PDOS) for the adsorbed molecule in configuration of Figure 3c. The energetic position of the corresponding orbitals of the free molecule is indicated by vertical lines of the corresponding color, the Fermi energy is at E=0. (d) Measured dI/dU spectrum on the lobe and at the Co-center compared to the spectrum on the clean CoO surface.

In Figure 5c we show the calculated projected density of states of the adsorbed molecules. Compared to the free molecule (energetic positions of the frontier orbitals indicated as lines in Fig. 5c) there is a slight charge transfer of $-0.2e$ (where $-e$ is the charge of an electron) from the molecule to the substrate.

The chosen U–J parameter ensured that the unoccupied part of the Co d-orbitals coincide energetically with the C and N orbitals of the molecular backbone as suggested by the experimental data of Fig. 5 a. This choice of U–J, however, does not influence the adsorption geometry of the molecule which can be easily understood since only 14% of the binding energy is due to the Co-O bond to the substrate. The resulting PDOS of the adsorbed molecule (Fig. 5 c) suggests that the HOMO of the molecule is situated only slightly below $E_F$. While the dominant C/N character of the HOMO agrees with experimental STM images the energetic position does not. This leads to a good agreement between the measured topography at $U_{bias}$ = -1 V and the simulated STM image at only $U_{bias}$ = -0.5 V (Fig. 5b) since at -1 V the Co orbital is dominating (see PDOS). Resolving the intramolecular electronic structure by STS also demonstrates that the HOMO has mainly C/N character and is located 0.8 eV below $E_F$ (Fig. 5 d). In consequence the DFT HOMO-LUMO gap of 1.5 eV underestimates the experimental gap of 2.1 eV. We argue that this is a failure of the employed approximations and would require post-DFT methods like GW to estimate the correct quasi-particle energies.

## 4. Conclusions

We demonstrated the detailed analysis of the adsorption properties of CoPc molecules on a thin layer of CoO on Ir(100). Experimentally we find well-defined adsorption geometries and attractive interactions between molecules that result in ordered compact or chain-like superstructures. Using state-of-the-art vdW-DFT calculations we showed that the naively expected covalent bond between the central Co atom of the molecule with an oxygen atom of the substrate exists but only contributes weakly (14%) to the total binding energy of 3.6 eV while the rest is van-der-Waals interaction. Of course, viewed as an energy *per atom* this is a sizable contribution but without vdW contribution the Co-O bond is not strong enough to bind the molecule on the cobalt oxide surface. In total the vdW interaction leads to a rather flat potential energy landscape (corrugation < 100 meV) even between adsorption sites at which the bonding O atom has to be pulled away from its original position in the oxide film by 0.9 Å. As judged from the vdW-DFT calculations other distortions of the substrate are weak, short-ranged and similar as those caused by a H2-Pc without the central metal-oxygen bond to the substrate. By calculating the interaction between molecules while ignoring the substrate we found that the experimentally determined intermolecular distances and orientations in the superstructure correspond very closely to the energetic minimum of the vdW potential energy. This indicates that also the lateral molecular interaction is dominated by dispersion forces.

We also determined the HOMO and LUMO orbitals experimentally by STS and compared them to the vdW-DFT calculations. While with a carefully chosen on-site Coulomb repulsion of the d-electrons on the central Co atom one improves the qualitative agreement of the character of the frontier orbitals, the reproduction of the experimental energetic positions of the orbitals needs post-DFT methods to calculate the true quasi-particle energies.

## Acknowledgements

Financial support from the Austrian Science Fund (Project No. 45, Functional Oxide Surfaces and Interfaces – FOXSI) and from the Deutsche Forschungsgemeinschaft (Research Unit FOR 1878 "funCOS") is acknowledged. We thank the Vienna Scientific Cluster (VSC) for ample supply of CPU time. The authors declare no competing financial interest.